\documentclass{ifacconf}

\usepackage{graphicx}      
\usepackage{natbib}        

\usepackage{xcolor}
\usepackage{url}
\usepackage{amsmath}
\usepackage{amssymb}
\usepackage{algorithm}
\usepackage{algpseudocodex}

\usepackage{mathtools}  
\usepackage{svg}

\usepackage{acronym}
\acrodef{OLS}{ordinary least squares estimator}
\acrodef{MLE}{maximum likelihood estimator}
\acrodef{SNR}{signal-to-noise ratio}
\acrodef{LTI}{linear time-invariant}
\acrodef{LQR}{linear quadratic regulator}
\acrodef{CE}{certainty equivalence}
\acrodef{PE}{persistency of excitation}
\acrodef{KL}{Kullback-Leibler}

\DeclareMathOperator*{\argmin}{arg\,min}
\newcommand{\R}{\mathbb{R}}

\newcommand{\Prob}{\mathbb{P}}
\newcommand{\Expect}{\mathbb{E}}
\newcommand{\N}{\mathcal{N}}
\newcommand{\SSet}{\mathcal{S}}

\newcommand{\bigO}{\mathcal{O}}
\newcommand{\simiid}{\stackrel{\text{i.i.d.}}{\sim}}

\newcommand{\tr}[1]{\mathrm{tr}\left({#1}\right)}

\begin{document}
\begin{frontmatter}

\title{Finite-Sample LTI System Identification from a Misspecified Finite Model Set\thanksref{footnoteinfo}}

\thanks[footnoteinfo]{
We acknowledge the support by Deutsche Forschungsgemeinschaft
(DFG, German Research Foundation) under Germany’s
Excellence Strategy - EXC 2075 – 390740016.
}

\author[First]{Nicolas Chatzikiriakos} 
\author[First]{Andrea Iannelli}

\address[First]{Institute for Systems Theory and Automatic Control, University of Stuttgart (e-mail: \{chatzikiriakos, iannnelli\}@ist.uni-stuttgart.de)}

\begin{abstract}                
    We consider the problem of finite-sample identification of an unknown LTI system from a finite hypothesis class. In contrast to existing results that assume the hypothesis class contains the true system, we consider the misspecified setting, where the true system does not belong to the hypothesis class. We first characterize the resulting estimation target and its dependence on the data-generation scheme. We then establish a finite-sample lower bound on the failure probability that any identification algorithm must incur. Finally, we analyze the robustness of an existing exponential weighting-based identification algorithm to the misspecification and derive a finite-sample upper bound on its failure probability. Our results illustrate when using a misspecified finite hypothesis class can provide an advantage over identification without prior knowledge of the candidate systems.
\end{abstract}


\end{frontmatter}

\section{Introduction}
Identifying an unknown dynamical system from a finite hypothesis class is a problem that appears naturally in many applications~\citep{Choirat2012}. In particular, the identification of the active mode of a switched system is a relevant problem in the adaptive control literature~\citep{narendra_adaptive_1997}. Recently, in the context of adaptive control \cite{rantzer_minimax_2021} considered model uncertainty in the form of a finite hypothesis class, renewing interest in this setting.
Furthermore, in evolutionary and ecological models often several hypothesis exist and the most suitable one needs to be determined from observations~\citep{johnson_model_2004}.
While all these applications have a finite hypothesis class, in many realistic cases the true system is not necessarily contained in the hypothesis class. 
Instead, the hypothesis class typically contains an approximation of the true system the learner should identify. 
However, recent works that cover the identification of a \ac{LTI} system from a finite hypothesis class rely on the assumption that the true system lies inside the hypothesis class and do not consider the misspecified setting. 
We remove this assumption and provide finite-sample identification guarantees. 

\textbf{Related Works:}
The problem of system identification has a rich history in the asymptotic regime~\citep{ljung_system_1999}. 
Based on novel tools in high-dimensional statistics~\citep{abbasi-yadkori_improved_2011, wainwright_high-dimensional_2019} recently this has been considered from a finite sample perspective (see, e.g., \citep{tsiamis_statistical_2023,ziemann_tutorial_2023} for an overview on the topic). 
In particular, the first line of works analyzed the estimation error of the \ac{OLS} in terms of sample complexity upper bounds \citep{simchowitz_learning_2018,sarkar_near_2019,dean_sample_2020} as well as sample complexity lower bounds~\citep{jedra_sample_2019,jedra_finite-time_2022}. 
Extending these works \citep{wagenmaker_active_2020,wagenmaker_task-optimal_2021,chatzikiriakos_hidden_2025} consider the problem of experiment design in this setting from a finite sample perspective.

More recently, the problem of identification from a finite hypothesis class has been considered for random excitations~\citep{chatzikiriakos_sample_2024,muehlebach_sample_2025} and in a more general setting where the excitation input is selected by the learner~\citep{chatzikiriakos_high_2026}.
These works, however rely on the assumption that hypothesis class contains the true system.

Recently, the topic of learning under misspecification has appeared in multiple contexts, e.g., \citep{xin2023learning,modi2024joint,lee_nonasymptotic_2024}. 
However, a finite sample analysis focussing on the interaction of data-generation and estimation target is missing. 
In addition, to the best of our knowledge the setting with a finite hypothesis class has only been considered for the case where the true system is contained in the hypothesis class. 

\textbf{Contribution:}
The contributions of this manuscript are summarized as follows
\begin{enumerate}
    \item We precisely characterize the \emph{estimation target} in a misspecified setting based on the \ac{KL}-information criterion. Further, we display its dependency on the data-generation process and provide a systems-theoretic interpretation. 
    \item We provide a failure probability lower bound for identification in the misspecified setting. By doing so we extend previous proof techniques to account for the fact that the true system is \emph{not} contained in the hypothesis class.   
    \item Extending previous work we provide a failure probability upper bound for an estimation algorithm based on exponential weighing in the misspecified setting. 
    Doing so we show that the algorithm is robust with respect to misspecifications in the hypothesis class. Further, we show that the algorithm almost achieves the fundamental limit indicated by our lower bound.  
    \item We discuss the effects of the results on identification and experiment design. In particular, we provide a corollary for the practically relevant case that a subset of the hypothesis class approximates the true dynamics with high accuracy and the other models differ significantly.       
\end{enumerate}
\par 
\textbf{Notation:}
Given a square matrix $M$ we denote its spectral radius and trace by $\rho(M)$ and $\tr{M}$, respectively.
We use $\mathbb{S}_{++}^{n_x}$ ($\mathbb{S}_{+}^{n_x}$)  to denote all symmetric positive definite (semi-definite) matrices of dimension $n_x \times n_x$. 
Given a matrix $Q \in \mathbb{S}_{++}^{n_x}$ we denote the generalized norm of a vector $x\in \R^{n_x}$ as $\Vert x \Vert_Q\coloneqq \sqrt{x^\top Q x}$. When $Q=I$ we write $\Vert x \Vert$.
Given matrices $Q, R \in \mathbb{S}_{+}^{n_x}$ we define 
$\Vert M \Vert_{Q, R} = \sqrt{\mathrm{vec}(M)^\top (R^\top \otimes Q) \mathrm{vec}(M)}$, where  $\otimes$ denotes the Kronecker product and $\mathrm{vec}(\cdot)$ is the vectorization operation.
When $Q = I$ we simplify notation and write $\Vert M \Vert_{R}$. 
When $Q=R=I$, the expression is equal to the Frobenius norm of $M$, hence we write $\Vert M \Vert_\mathrm{F}$.
We use the shorthand $\Prob_\theta[\cdot]$ ($\Expect_\theta[\cdot]$) when referring to the probability (expectation) given that the system generating the data is $\theta$.
When this is clear from the context or there are multiple sources of randomness we drop the subscript for clarity.  
Given a set $\omega \in \Omega$ we denote its complement by $\omega^c \coloneqq \Omega \setminus\omega$.
\section{Problem Setting}
Consider an unknown \ac{LTI} system
\begin{equation}
    x(t+1) = A_* x(t) + B_* u(t) + w(t),\label{eq:TrueSysEvo}
\end{equation}
where $x(t) \in \R^{n_x}$, $u(t)\in \R^{n_u}$ are the state and the input of the system at timestep $t$. To simplify notation we define $z(t) \coloneqq \begin{bmatrix}x(t) ^\top & u(t)^\top \end{bmatrix}^\top \in \R^{n_z}$. The process noise is denoted by $w(t)$. We assume $w(t) \stackrel{\text{i.i.d}}{\sim} \mathcal{N}(0, \sigma_w^2 {I}_{n_x})$ and consider an identification problem where the learner has access to a finite hypothesis class
\begin{equation} \label{eq:defSet}
    \SSet \coloneqq \{(A_0, B_0), \dots, (A_{N_\mathrm{sys}}, B_{N_\mathrm{sys}})\}, 
\end{equation} 
where set $\SSet$ might not contain the true system.
The goal of the learner is to identify a system $\theta_i \coloneqq  (A_i, B_i)\in \SSet$ that is closest to $\theta_*$ using the data $\mathcal{D}_T  = \{x(t), u(t)\}_{t=0}^{T-1}$.
Unless specified otherwise we consider the case where $\theta_* \coloneqq (A_*, B_*)  \notin \SSet$. 
For the remainder of this work we impose the following assumption on the data collection.
\begin{assum}
During data collection the input is sampled according to $u(t) \simiid \N(0, \Sigma_u)$.
\end{assum}
\begin{rem}
We do not further restrict the data collection scheme and present a framework that can cover i.i.d. data ($\Expect[z(t) z(t)^\top] = \Sigma_z,\,\forall t\ge 0$), single trajectory data, and data consisting of multiple independent trajectories.
\end{rem}
Since the learner has only access to one realization of the random data, we characterize \emph{closeness} of the candidate systems from an information theoretic perspective.
To formalize this, denote the probability of observing the data $\mathcal{D}_T$ from system $\theta_i$ as $f_{\theta_i}(\mathcal{D}_T)$. 
Since the data $\mathcal{D}_T$ is randomly drawn from a data distribution  we want the learner to pick the estimate $\hat \theta_*$ that minimizes the 
relative entropy or KL-divergence \citep{cover1991elements} with respect to $\theta_*$ under the data distribution, i.e., the goal of identification, or estimation target, is
\begin{equation}\label{eq:KL-InformationCriterion}
   \hat \theta_* \in \argmin_{\theta \in \SSet} \mathrm{KL}(f_{\theta_*}(\mathcal{D}_T) \Vert f_{\theta}(\mathcal{D}_T )). 
\end{equation}
The objective in~\eqref{eq:KL-InformationCriterion} is known as the Kullback-Leibler information criterion~\citep{white1982Misspecified}.
Informally,~\eqref{eq:KL-InformationCriterion} yields the estimate $\hat \theta_*$ that minimizes the \emph{distance}\footnote{Note that the KL-divergence is not a distance metric and does not satisfy the triangle inequality. Nevertheless, this thinking can be instructive to build intuition.}
between $f_{\theta_*}(\mathcal{D}_T)$ and $f_{\hat \theta_*}(\mathcal{D}_T)$ and hence picks the estimate $\hat \theta_*$ that is closest to $\theta_*$ for the data collection scheme in terms of the KL-divergence. 
\begin{prop}
    The set of optimizers of the KL information criterion~\eqref{eq:KL-InformationCriterion}, i.e., the estimation target, is equivalent to 
    \begin{equation}\label{eq:estTarget}
        \hat\Theta_{*,T} \coloneqq \argmin_{\theta\in \SSet} \Vert \theta_* - \theta \Vert^2_{\bar \Sigma_z(T)}
    \end{equation}
    where $\bar \Sigma_z(T)$ is the average covariance over $T$, i.e.,  
\begin{equation}
    \begin{aligned}
        \bar \Sigma_z(T) &\coloneqq \frac{1}{T} \sum_{t=0}^{T-1} \Expect \left[z(t) z(t)^\top \right] \\
        &= \begin{bmatrix}
        \frac{1}{T}  \sum_{t=0}^{T-1} \Expect\left[x(t)x(t)^\top\right] & 0 \\ 0 &\Sigma_u 
    \end{bmatrix}.    
    \end{aligned}
\end{equation}
\end{prop}
\begin{pf}
    Using the definition of the KL-divergence yields
    \begin{align*}
        \mathrm{KL}&(f_{\theta_*}(x_{1:T}) \Vert f_{\theta}(x_{1:T} )) = \Expect_{\theta_*} \left[\log\left(\frac{f_{\theta_*}(\mathcal{D}_T)} {f_{\theta}(\mathcal{D}_T)}\right)\right]\\
        & = \Expect_{\theta_*}\left[\frac{\eta}{\sigma_w^2} \left(\sum_{t=0}^{T-1}\Vert x(t+1) - \theta z(t)\Vert^2 - \Vert w(t)\Vert^2\right)\right] \\
        &= \Expect_{\theta_*}\left[\frac{\eta}{\sigma_w^2} \Big(\sum_{t=0}^{T-1}\Vert (\theta_*\!-\!\theta) z(t) \Vert^2\!-\!2 w(t)^\top (\theta_*\! -\!\theta)z(t) \Big)\right]\\
        &= \frac{\eta}{\sigma_w^2} \Expect_{\theta_*}\left[\tr{(\theta_*- \theta)^\top (\theta_*- \theta) \sum_{t=0}^{T-1} z(t)z(t)^\top}\right]
    \end{align*}
    where we used that $w(t)$ is i.i.d. zero-mean, and leveraged the cyclic property of the trace. Using the decomposition $\bar \Sigma_z(T) = (\bar \Sigma^{\frac{1}{2}}_z)^\top \bar \Sigma^{\frac{1}{2}}_z$ and elementary reformulations yields
    \begin{align*}
        \frac{\eta}{\sigma_w^2} &\Expect_{\theta_*}\left[\tr{(\theta_*- \theta)^\top (\theta_*- \theta) \sum_{t=0}^{T-1} z(t)z(t)^\top}\right] \\
        &= \frac{T \eta}{\sigma_w^2} \tr{\bar \Sigma^{\frac{1}{2}}_z (\theta_*- \theta)^\top (\theta_*- \theta)(\bar \Sigma^{\frac{1}{2}}_z)^\top } \\
        &= \frac{T \eta}{\sigma_w^2} \tr{((\theta_*- \theta)(\bar \Sigma^{\frac{1}{2}}_z)^\top)^\top (\theta_*- \theta)(\bar \Sigma^{\frac{1}{2}}_z)^\top }\\
        &= \frac{T \eta}{\sigma_w^2} \Vert \theta_*-\theta\Vert^2_{\bar \Sigma_z(T)}.
    \end{align*}
    The result follows since constant factors don't change the optimizers.
\end{pf}
It is important to realize that the dependence of $\hat \Theta_{*,T}$ on the data collection scheme is natural in the misspecified setting. 
Since the learner does not have access to $(A_*, B_*)$ it can only judge similarity of a candidate $(A_i, B_i)$ to $(A_*, B_*)$ through the observed data. The data distribution however might highlight the differences in some directions more than others leading to the weighting in~\eqref{eq:estTarget}.
\begin{rem}
When $\theta_* \in \SSet$ and \ac{PE} is satisfied (cf. \cite[Definition 2]{chatzikiriakos_high_2026}) it follows directly that $\hat\theta_* = \theta_*$, since for two probability distributions $p$ and $q$ it holds that $\mathrm{KL}(p\Vert q) = 0 \Leftrightarrow p=q$.
In this case the optimal estimate is independent of the data generation, whereas in the misspecified setting the data collection scheme generally influences the estimation target $\hat \theta_*$ defined in~\eqref{eq:KL-InformationCriterion}.
\end{rem}
We seek an estimator that with high probability identifies an element of the set $\hat \Theta_{*, T}$, which depends on the \emph{data distribution} while only observing a \emph{realization} of this distribution.


\section{Sample Complexity Under Misspecification}
In this section, we first introduce $c_{\theta_i}^\mathrm{dist}$-KL distinguishability, a core property allowing identification of $\hat \Theta_{*,T}$. 
We then present sample complexity lower bounds revealing fundamental limits in the identification from a finite hypothesis class. 
Finally, analyze an estimation algorithm that was shown to be optimal when $\theta_* \in \SSet$ in terms of its failure probability in the misspecified setting.    
Throughout we consider a fixed budget perspective, i.e., given a fixed budget of samples $T$ we aim to identify the true system with a possibly low failure probability $\delta$. 
Similar result can be obtained in the fixed confidence setting by using an algorithm like \cite[Algorithm 1]{chatzikiriakos_high_2026} and adapting our analysis slightly.    
To simplify notation we define $\Delta \theta_{ji} \coloneqq \theta_j  - \theta_i$. When taking $\theta_*$ as the baseline we write $\Delta \theta_i \coloneqq \theta_* - \theta_i$. Overloading notation, we write $\Delta \hat \theta_i \coloneqq \hat \theta_* - \theta_i$ and  $ \Delta \hat \theta_* \coloneqq \theta_* - \hat \theta_*$. 
\subsection{KL distinguishability}
\begin{defn}[$c_{\theta_i}^\mathrm{dist}$-KL distinguishability]\label{def:PE_extended}
    We say 
    $\hat \Theta_{*, T}$
    is  $c_{\theta_i}^\mathrm{dist}$-KL distinguishable from $\hat \Theta_*^c \coloneqq \SSet\setminus \hat \Theta_{*, T}$ under $(\theta_*, \bar \Sigma_z(T))$
    if there exist coefficients $\{c_{\theta_i}^\mathrm{dist}\}_{\theta_i \in \hat \Theta_{*,T}^c}\in (0, 1)$ such that
    for any $x(0) \in \R^{n_x}$, and $\theta_i \in \hat \Theta_{*,T}^c$
    \begin{equation}\label{eq:KL_block_dist_1}
        \sum_{t=0}^{T-1} \Expect \left[\Vert \Delta \theta_i z(t) \Vert^2 \right] -  \sum_{t=0}^{T-1} \Expect \left[\Vert \Delta \hat \theta_*z(t) \Vert^2 \right] \ge c_{\theta_i}^\mathrm{dist} c^\mathrm{PE}_{\theta_i} T ,
    \end{equation} 
    where the constants $c^\mathrm{PE}_{\theta_i} >0$ are defined such that
    \begin{equation}\label{eq:KL_block_dist_2}
        \sum_{t=0}^{T-1} \Expect \left[\Vert \Delta \theta_i z(t) \Vert^2 \right]  \ge c^\mathrm{PE}_{\theta_i} T >0 \quad \forall \theta_i \in  \hat \Theta_{*,T}^c. 
    \end{equation}  
\end{defn}
Observe that Definition~\ref{def:PE_extended} requires $c_{\theta_i}^\mathrm{PE}>0$ only for $\theta_i \in \hat \Theta_{*, T}^c$. Thus, our results cover also the case where $\theta_* \in \SSet$ but the excitation is insufficient for identification of $\theta_*$.
Note that there always exists a coefficient $c_{\theta_i}^\mathrm{dist} >0$ so that Definition~\ref{def:PE_extended} is satisfied. In the following, we will show that this is a key quantity determining the achieved failure probability.
\begin{rem}\label{rem:distinguishabilityPE}
When $\theta_* \in \SSet$ and the data is collected such that $\theta_*= \hat \Theta_{*, T}$, Definition~\ref{def:PE_extended} is identical to the definition of \ac{PE} for this setting (cf. \cite[Definition 2]{chatzikiriakos_high_2026}). 
\end{rem}
\subsection{Failure probability lower bounds}
To investigate fundamental limits we consider a lower bound that holds for \emph{any} reasonable algorithm. For this we slightly extend the definition of $\delta$-correct algorithms as in \citep{chatzikiriakos_high_2026}.
\begin{defn}
    Consider the setup described in Section 2. An algorithm is called $\delta$-correct, if for any $T > 0$ and any $(\theta_*, \SSet)$ there exists a $\delta(T) \in (0, 1)$ such that the algorithm
returns an estimate $\hat \theta_{T}$ that satisfies 
\begin{equation}\label{eq:delta-stable}
    \Prob\left[\hat \theta_{T} \in \hat \Theta_{T, *}\right] \ge 1-\delta(T).
\end{equation}
\end{defn}
Note that given a desired failure probability $\bar \delta$ 
\begin{equation}
    \bar T(\bar\delta) = \min T \text{ s.t. \eqref{eq:delta-stable} holds with }  \delta(T) \le \bar \delta  
\end{equation}
is the sample complexity of a $\delta$-correct algorithm.
The following theorem provides a lower bound on the failure probability any $\delta$-correct algorithm needs to satisfy. 
\begin{thm}\label{th:LowerBound}
    Consider the unknown system \eqref{eq:TrueSysEvo}, hypothesis class~\eqref{eq:defSet} and estimation target~\eqref{eq:estTarget}. Then given data of length $T$ any $\delta$-correct algorithm satisfies
    \begin{equation*}
        \begin{aligned}
            \delta(T) \ge \max_{\theta_i \in \hat \Theta_{T, *}^c}\frac{1}{2.4}e^{-\frac{T}{2 \sigma_w^2 c_{\theta_i}^\mathrm{dist}}  \mathrm{tr}\left((\Delta \theta_i - \Delta \hat \theta_*)^\top(\Delta \theta_i - \Delta \hat \theta_*) \bar \Sigma_z(T) \right) }
        \end{aligned}
    \end{equation*}
\end{thm}
\begin{pf}
    Consider the log-likelihood ratio
    \begin{equation}
        L_t = \log\left(\frac{f_{\hat \theta_*}(z(0), \dots, z(T-1))}{f_{\theta_i}(z(0), \dots, z(T-1))}\right). 
    \end{equation} 
    Performing the same steps as in \cite{jedra_sample_2019} we obtain 
    \begin{equation*}
        \Expect_{\theta_*}[L_T] = \Expect_{\theta_*}\left[\sum_{s=0}^{T-1}\Expect_{\theta_*}\left[\log\left(\frac{f_{\hat \theta_*}(x(s+1)\vert z(s))}{f_{\theta_i}(x(s+1)\vert z(s))}\right)\vert \mathcal{F}_{s}\right]\right]. 
    \end{equation*}
    Note that, conditioned on $\mathcal{F}_s$, $f_{\theta}(x(s+1)\vert z(s))$ is a Gaussian with mean $\mu = \theta z(s-1)$ and covariance $\sigma_w^2 I_{n_x}$. 
    However, since the expectation is taken with respect to $\theta_* \notin \SSet$ we cannot leverage the definition of the KL-divergence of two Gaussians directly.
    Instead, using the definition of multivariate Gaussians, we obtain 
    \begin{equation}
        \log\left(\frac{f_{\hat \theta_*}(x\vert z)}{f_{\theta_i}(x\vert z)}\right) = \frac{1}{2\sigma_w^2}(\Vert x - \theta_i z\Vert^2 - \Vert x - \hat \theta_* z\Vert^2).  
    \end{equation}    
    Using that $x(s+1)=\theta_* z(s) +w(s)$, the definition of the norm and taking the expectation for both terms individually we obtain 
    \begin{equation}
        \begin{aligned}
                    \Expect_{\theta_*}&\left[\log\left(\frac{f_{\hat \theta_*}(x(s+1)\vert z(s))}{f_{\theta_i}(x(s+1)\vert z(s))}\right)\vert \mathcal{F}_{s}\right] \\ 
                    &\qquad = \frac{1}{2\sigma_w^2} \left(\Vert\Delta \theta_i z(s) \Vert^2 -  \Vert  \Delta \hat \theta_* z(s) \Vert^2\right). 
        \end{aligned}
    \end{equation}
    Using the cyclic property of the trace similar as in~\cite{jedra_sample_2019} we obtain 
        \begin{align}
        &\Expect_{\theta_*}[L_T] \\ &= \frac{1}{2\sigma_w^2} \mathrm{tr}\Big((\Delta \theta_i\! - \!\Delta \hat \theta_*)^\top \! (\Delta \theta_i \!-\! \Delta \hat \theta_*)\!\sum_{s=0}^{T-1} \Expect_{\theta_*}\left[z(s) z(s)^\top \right]\Big). \notag 
        \end{align}
    Further, we have that 
    \begin{equation}
        \begin{aligned}
            \Expect_{\theta_*}[L_T] &= \Expect_{\theta_*} \left[\log\left(\frac{f_{\hat \theta_*}}{f_{\theta_i}}\right) + \log\left(\frac{f_{\theta_*}}{f_{\theta_*}}\right)\right] \\
            &= \Expect_{\theta_*} \left[\log\left(\frac{f_{\theta_*}}{f_{\theta_i}}\right)\right] - \Expect_{\theta_*} \left[\log\left(\frac{f_{\theta_*}}{f_{\hat \theta_*}}\right)\right] \\
            &= \mathrm{KL}(f_{\theta_*} \Vert f_{\theta_i}) - \mathrm{KL}(f_{\theta_*} \Vert f_{\hat \theta_*}).
        \end{aligned}
    \end{equation}
    Using the definition of $c_{\theta_i}^\mathrm{dist}$ we have 
    \begin{equation}
        \mathrm{KL}(f_{\theta_*} \Vert f_{\theta_i}) - \mathrm{KL}(f_{\theta_*} \Vert f_{\hat \theta_*}) \ge c_{\theta_i}^\mathrm{dist} \mathrm{KL}(f_{\theta_*} \Vert f_{\theta_i}). 
    \end{equation} 
    From here we can use the data-processing inequality \citep{garivier_explore_2019} to obtain that 
    \begin{equation}
         \Expect_{\theta_*}[L_t] \ge c_{\theta_i}^\mathrm{dist} \mathrm{kl}(\Prob_{\theta_*}(\mathcal{E})\Vert \Prob_{\theta_i}(\mathcal{E})),
    \end{equation}
    where $\mathrm{kl}(p\Vert q)$ is the KL-divergence of two Bernoulli distributions.
    With the event $\mathcal{E} = \{\hat \theta_T \in \hat \Theta_{*,T}\}$ we have
    \begin{align*}
        \Prob_{\theta_*}(\mathcal{E}) &\ge 1-\delta \\
         \Prob_{\theta_i}(\mathcal{E}) &\le \delta
    \end{align*}   
    since we consider $\delta$-correct algorithms.
    Following the steps in \cite{chatzikiriakos_sample_2024} we then obtain that any $\delta$ correct algorithm satisfies 
    \begin{equation*}
        \begin{aligned}
            \mathrm{tr}\Big((\Delta \theta_i - \Delta \hat \theta_*) ^\top (\Delta \theta_i - \Delta \hat \theta_*) & \sum_{s=0}^{T-1} \Expect_{\theta_*}\left[z(s) z(s)^\top \right]\Big) \\ 
            & \quad \ge 2 \sigma_w^2 c_{\theta_i}^\mathrm{dist} \log\left(\frac{1}{2.4 \delta}\right).
        \end{aligned}
    \end{equation*}
    Solving the inequality for $\delta$ yields the result. 
\end{pf}
Theorem~\ref{th:LowerBound} provides insights into the hardness of learning under misspecification. 
In particular, we observe that the relative distance between a candidate $\theta_i$ and $\hat\theta_*$ measured by the respective KL-divergences w.r.t. the ground truth system $\theta_*$ governs how fast the failure probability decays. 
Note that Theorem~\ref{th:LowerBound} lacks a dimensional dependency on the hypothesis class. 
We suspect that such a dependency is necessary, indicating that the lower bound can be improved, e.g., by adapting the argument in the proof of \cite[Theorem~2]{jedra_finite-time_2022}.

\subsection{Failure probability upper bound}
We employ an exponential weighting-based estimation scheme that has been shown to be optimal in the setting where the true system is contained in the hypothesis class~\citep{chatzikiriakos_high_2026}. 
In the following, we investigate the robustness of this estimator to a misspecified hypothesis class. 
To this end, we define the empirical prediction errors
\begin{equation}\label{eq:defPrederror}
    \varepsilon_{\theta_i}(T)\coloneqq \sum_{t=0}^{T-1} \Vert x(t) - \theta_i z(t-1) \Vert^2 \quad \forall \theta_i \in \SSet. 
\end{equation}
Then we build an exponential weighting 
\begin{equation}\label{eq:exp_weights}
    w_T(\theta_i) \propto e^{-\eta \varepsilon_{\theta_i}(T)}, 
\end{equation}
where $\eta$ is a learning rate, and we normalize the weights $w_{\theta_i}$ such that $\sum_{i=0}^{N_\mathrm{sys}} w_{T}(\theta_i) = 1$. 
Note that, for appropriate choice of $\eta$ the weights~\eqref{eq:exp_weights} coincide with the Gaussian posterior of $\theta_i$ given the data.
With this, our estimate is drawn from this posterior over candidate system, i.e., 
\begin{equation} \label{eq:estimator}
    \hat \theta_T \sim w_T(\theta_i).
\end{equation}
The following theorem provides a sample complexity guarantee for the estimator~\eqref{eq:estimator}.
\begin{thm}\label{th:UpperBound}
Consider the unknown system~\eqref{eq:TrueSysEvo} and the set~\eqref{eq:defSet}.
Then, for the estimator~\eqref{eq:estimator} we have  
\begin{equation} \label{eq:high-prob-ID}
    \Prob\left[\hat \theta_T \in \hat\Theta_{*,T}\right] \ge 1 - \sum_{\theta_i \in \hat \Theta_{*,T}^c} e^{-\frac{\eta}{2} T c_{\theta_i}^\mathrm{dist} c_{\theta_i}^\mathrm{PE}},
\end{equation}
if $\eta < \min_{\theta_i \in \hat \Theta_{*, T}^c} \frac{c_{\theta_i}^\mathrm{dist} c^\mathrm{PE}_{\theta_i}}{4 n_z \bar \rho_{\theta_i}^2}$ where $\bar\rho_{\theta_i}$ depends on the data collection scheme and is defined in~\eqref{eq:barRho_traj} and~\eqref{eq:barRho_iid}.
\end{thm}
\begin{pf}
    Note that for any $\hat \theta_* \in \hat \Theta_{*, T}$ we have
    \begin{equation}
        \Prob[\hat \theta_T = \theta_i] \le \Expect\left[e^{-\eta \left(\varepsilon_{\theta_i}(T) - \varepsilon_{\hat \theta_*}(T) \right)}\right]
    \end{equation}
    by following the reasoning in \cite[Lemma C.1]{muehlebach_sample_2025}. 
    Further, we have
        \begin{align}
            \varepsilon_{\theta_i}&(T) - \varepsilon_{\hat \theta_*}(T)\label{eq:decompositionError}\\ 
            &\le \sum_{t=0}^{T-1}\Vert \Delta \theta_i z(t)\Vert^2 - \Vert \Delta \hat \theta_* z(t)\Vert^2 + 2 w(t)^\top \Delta \hat \theta_{i} z(t). \notag
        \end{align}
    Note that by Proposition~\ref{prop:derivXi} we can rewrite \eqref{eq:decompositionError} into the form 
    \begin{equation}
        \varepsilon_{\theta_i}(T) - \varepsilon_{\hat \theta_*}(T) = \zeta^\top \Xi_{\theta_i} \zeta,
    \end{equation} 
    where $\zeta\sim \N(0, I)$ and $\Xi_{\theta_i} = \Xi_{\theta_i}^\top$ depends on the data collection scheme.
    Note that \begin{equation}
        c_{\theta_i}^\mathrm{dist}c_{\theta_i}^\mathrm{PE} T \le \tr{\Xi_{\theta_i}} \le T n_z \bar \rho_{\theta_i}
    \end{equation}
    and hence $ c_{\theta_i}^\mathrm{dist}c_{\theta_i}^\mathrm{PE} \le n_z \bar \rho_{\theta_i}$. Thus by the condition on $\eta$  
    \begin{equation}
        \eta \le \frac{ c_{\theta_i}^\mathrm{dist}c_{\theta_i}^\mathrm{PE}}{4 n_z \bar \rho_{\theta_i}^2} \le \frac{1}{4 \bar \rho_{\theta_i}} \le \frac{1}{4 \rho(\Xi_{\theta_i})}.
    \end{equation} 
   Hence, we can apply Proposition~\ref{prop:SubExp_new} to obtain 
    \begin{equation}
        \Prob[\hat \theta_T = \theta_i] \le \exp\left(-\eta \left(\tr {\Xi_{\theta_i}}-2\eta\Vert \Xi_{\theta_i}\Vert_\mathrm{F}^2\right)\right).
    \end{equation} 
    Invoking the condition on $\eta$ and leveraging the statements of Proposition~\ref{prop:derivXi} we obtain 
    \begin{equation}
     \Prob[\hat \theta_T = \theta_i] \le e^{-\frac{1}{2}\eta \tr {\Xi_{\theta_i}}}.
    \end{equation} 
    Since by definition we have 
    \begin{equation}
        \tr{\Xi_{\theta_i}} = \sum_{t=0}^{T-1}\Expect\big[\Vert\Delta\theta_i z(t)\Vert^2\big]-\Expect\big[\Vert\Delta\hat\theta_* z(t)\Vert^2\big]
    \end{equation} 
    we can then apply Definition~\ref{def:PE_extended} and apply a union bound to conclude the proof.
\end{pf}

Comparing Theorems~\ref{th:LowerBound} and~\ref{th:UpperBound} we observe that both results almost match in terms of the key problem specific factors. However we observe that Theorem~\ref{th:UpperBound} exhibts a dependency on $\vert \Theta_{*,T}^c\vert$, i.e., the number of systems in $\SSet$ that need to be distinguished from $\hat \Theta_{*,T}$. As discussed after Theorem~\ref{th:LowerBound} we suspect that such a dependence is fundamental.
Secondly, the lower bound has an additional $(c_{\theta_i}^\mathrm{dist})^{-1}$ factor compared to the upper bound. This results in a gap between upper and lower bound that increases as $c_{\theta_i}^\mathrm{dist}$ decreases. 
Note that we suspect that this is an artifact of our proof techniques for Theorem~\ref{th:LowerBound}, since smaller $c_{\theta_i}^\mathrm{dist}$ increases the difficulty of learning. 
It is interesting problem for future research to investigate how the lower bound can be improved to match the upper bound.
Note that the choice of $\eta$ in Theorem~\ref{th:UpperBound} depends on potentially unknown quantities. 
This is due to the fact that the predictions errors scale with the level of excitation which depends on oracle knowledge such as the true system. 
We remark that a (potentially conservative) approximation of the upper bound on $\eta$ can be found using knowledge of the hypothesis class $\SSet$. 
For future work it could be interesting to investigate whether a regularization like in \cite{muehlebach_sample_2025} could remove this limitation. 

When using trajectory data, $\eta$ decreases with $A_*$ approaching instability or as $A_*$ becomes unstable (cf. \eqref{eq:barRho_traj}). This is due to the fact that we cannot resort to blocking techniques (cf., e.g., \cite{simchowitz_learning_2018}) in Proposition~\ref{prop:SubExp_new} since the matrix $\Xi$ is in general indefinite. Thus, a non-zero mean could not simply be omitted in Proposition~\ref{prop:SubExp_new}, which is true for the case where $\theta_* \in \SSet$ and enables the use of blocking techniques (in this case a non-zero mean can only improve the bound \eqref{eq:subExp_new}). Note that \cite{sarkar_near_2019} have shown that \ac{OLS} is statistically inconsistent for unstable systems. On the other hand, when $\theta_* \in \SSet$ \cite{chatzikiriakos_sample_2024} have shown that identification is possible even for unstable systems for a finite hypothesis class. It is an interesting direction to investigate whether this dependency of $\eta$ on the stability of $A_*$ can be circumvented. 
\section{Discussion of the results} \label{sec:discussion}
In the following we discuss multiple interesting consequences of our results.
\subsection{The benefit of using a finite hypothesis class.}
A natural question is whether it is beneficial to use knowledge of the misspecified hypothesis class or employ an \ac{OLS} estimator instead. 
In the following we consider a special case that allows for an answer to this question.  
\begin{cor}\label{co:Upper}
    Consider the same setup as in Theorem~\ref{th:UpperBound}. Further, assume that for any $\hat \theta_* \in \hat \Theta_{*, T}$ and any $\theta_i \in \hat \Theta_{*,T}^c$ there exists $\gamma >1$ and $\varepsilon>0$ so that 
    \begin{equation}
        \lambda_\mathrm{max}(\Delta \hat\theta_*^\top \Delta \hat\theta_*) \le \varepsilon^2 \quad \text{and} \quad \lambda_\mathrm{min}(\Delta \theta_i^\top \Delta \theta_i) > \gamma \varepsilon^2 .
    \end{equation}
    Select $\eta$ according to Theorem~\ref{th:UpperBound}. Then the estimator~\eqref{eq:estimator} satisfies
    \begin{equation}
     \Prob\left[\hat \theta_T \in \hat\Theta_{*,T}\right] \ge 1 - \sum_{\theta_i \in \hat \Theta_{*,T}^c} e^{-\frac{\eta}{2} T (\gamma-1)\varepsilon^2 \bar \Sigma_z(T)}.   
    \end{equation}
\end{cor}
The case in Corollary~\ref{co:Upper} describes a scenario where one model (or multiple) capture the true system dynamics accurately, whereas the others show a significant difference. 
In this case, we have
\begin{equation*}
    \begin{aligned}
    \sum_{t=0}^{T-1} \Expect \left[\Vert \Delta \theta_i z(t) \Vert^2 \right] &-  \sum_{t=0}^{T-1} \Expect \left[\Vert \Delta \hat \theta_*z(t) \Vert^2 \right] \\ &\ge (\gamma-1) \varepsilon^2 \Expect\left[\tr{\sum_{t=0}^{T-1} z(t) z(t)^\top}\right].
    \end{aligned}
\end{equation*}
This setting is especially important for many practical applications where the hypothesis class is build from first principles, e.g., fault detection.
Corollary~\ref{co:Upper} states that whenever $\gamma$ is large, this aids our algorithm in identification.       
Note that for the corresponding bounds on the \ac{OLS} if we fix $\varepsilon$ and $T$ we have $\delta = \bigO(e^{-\varepsilon^2 T})$ which is the same rate as in Corollary~\ref{co:Upper} (cf. \cite[Equation (16)]{tsiamis_statistical_2023}). However, if $\gamma \gg 1$ leveraging a finite hypothesis class can yield significant benefit compared to using \ac{OLS} for estimation.  
This effect is particularly significant in small data regimes. On the other hand, when a lot of data is available, using \ac{OLS} for estimation can result in a closer approximation, whereas the methods in this manuscript can only decrease the failure probability $\delta$. 
A numerical investigation of this aspect is provided in Section~\ref{sec:numerics}.
We see it as an interesting direction for future research to capture this tradeoff more precisely and investigate algorithms that can achieve the best of both worlds. 
%
%
\subsection{Influence of $\bar \Sigma_z(T)$ on the estimation target.}
The definition of $\hat\Theta_{*,T}$~\eqref{eq:estTarget} crucially depends on the data covariance $\bar \Sigma_z(T)$.  
When the true system is contained in the hypothesis class, experiment design can be used to decrease the failure probability of identification for a fixed number of samples~\citep{chatzikiriakos_high_2026}. In this setting the data generation scheme carries more significance since it also determines the estimation target. 
This is a natural consequence of the misspecification and opens multiple research directions. One interesting question is how robust the estimation target is under distribution shifts. 
This is of particular importance for applications in experiment design and online learning where the true system is often operated in different regimes.
\subsection{Modifying the estimation target trough reweighting.}
To allow for additional flexibility in the estimation we slightly generalize the estimation target~\eqref{eq:estTarget} by a weighting matrix $Q\succeq 0$ and define 
\begin{equation}
    \hat\Theta_{*,T}^{(Q)}  \coloneqq \{ \theta \in \argmin_{\theta\in \SSet} \Vert \theta_* - \theta \Vert_{{Q, \bar \Sigma_z(T)}}\}.
\end{equation}
With this we seek an estimator for which, given the data $\mathcal{D}_T$, there exists $\delta(T)$, so that 
\begin{equation}
    \Prob\left[\hat \theta_T \notin \hat \Theta_{*,T}^{(Q)} \right] \le \delta(T).
\end{equation}
Following the steps of the proof a similar result to Theorem~\ref{th:UpperBound} can be shown when adding the weighting to the estimator~\eqref{eq:defPrederror} and the definition of $c_{\theta_i}^\mathrm{dist}$-KL distingishability \eqref{eq:KL_block_dist_1}, \eqref{eq:KL_block_dist_2}.
Introducing such a weighting matrix $Q$ could be particularly useful when it is hard to influence $\Sigma_z(t)$.
\section{Numerical evaluation} \label{sec:numerics}
\begin{figure*}[t]
    \centering
    \includegraphics[width=0.7 \linewidth]{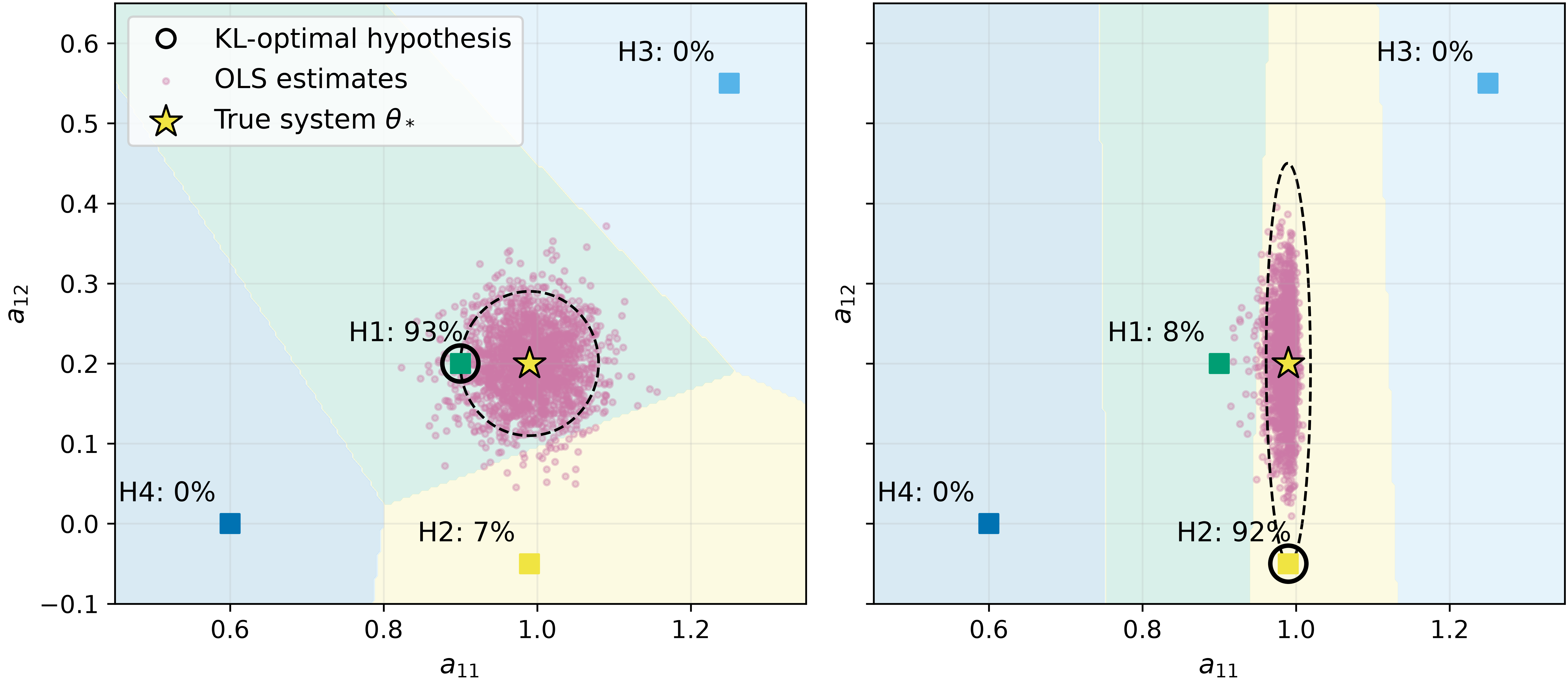}
    \caption{Candidate systems, empirical estimates of our algorithm and \ac{OLS} estimates over 2000 Monte-Carlo simulations for i.i.d data (left) and trajectory data (right). Shaded regions indicate which hypothesis would be selected when projecting the \ac{OLS} estimate on the closest candidate in the respective norm.}
    \label{fig:geometry}
\end{figure*}

To illustrate the theoretical findings we consider an unknown dynamical system of the form~\eqref{eq:TrueSysEvo} with 
\begin{equation} \label{eq:numericsTheta_Star}
A_* = \begin{bmatrix}
    0.99 & 0.2\\
    0. & 0.1
\end{bmatrix} \qquad \text{and} \qquad B_* = \begin{bmatrix}
    0.5 \\ 0.2
\end{bmatrix}.
\end{equation} 
Furthermore, we consider the hypothesis class $\SSet = \{(A_i, B_*)\}_{i=1}^4$ with 
\begin{equation}
    \begin{aligned}
        A_1 &= \begin{bmatrix}
        0.9 & 0.2\\
        0. & 0.1
    \end{bmatrix},
    \qquad  
    A_2 = \begin{bmatrix}
        0.99 & -0.05\\
        0. & 0.1
    \end{bmatrix}\\ 
    A_3 &= \begin{bmatrix}
        1.25 & 0.55\\
        0. & 0.1
    \end{bmatrix}
    \qquad 
    A_4 = \begin{bmatrix}
        0.6 & 0\\
        0. & 0.1
    \end{bmatrix}.
    \end{aligned}  
\end{equation}
In our simulations\footnote{The Python code used to reproduce the numerical results is available at \url{https://github.com/col-tasas/2026-misspecified-identification}} we consider two different data-generation scenarios. 
First, we collect i.i.d. data with isotropic sampling of $x(t), u(t) \simiid \N(0, I)$. 
Second, we consider the case of trajectory data where $u(t)\simiid \N(0, I)$.
For the numerical analysis we consider the estimates after collecting 250 samples over 2000 Monte-Carlo simulations for the estimator~\eqref{eq:estimator} and \ac{OLS}. The results are depicted in Figure~\ref{fig:geometry}.
%
%
It can be observed how the KL-optimal hypothesis $\hat \Theta_{T,*}$ in~\eqref{eq:estTarget}, which is a singleton in both cases, depends on the data generation. For i.i.d. samples, $\hat \Theta_{T,*}$ corresponds to the candidate systems $(A_1, B_*)$ which is closest to $(A_*, B_*)$ given the data-distribution. 
When considering trajectory data, $\hat \Theta_{T,*}$ corresponds to $(A_2, B_2)$.
Intuitively, this is due to the fact that \eqref{eq:numericsTheta_Star} can easily be excited in $x_1$ direction and much less in $x_2$ direction. 
This puts larger weight on the differences in $a_{11}$ compared to the differences in $a_{12}$. 
Importantly, this is not only a characteristic of the estimator. It can be seen that also the \ac{OLS} is sensitive to these different levels of excitation, which is in accordance with existing results on data-based, ellisoidal, identification error bounds for \ac{OLS} (see, e.g., \citep{dean_sample_2020,chatzikiriakos_end--end_2026}).

\begin{figure}[h]
    \centering
    \includegraphics[width=0.99\linewidth]{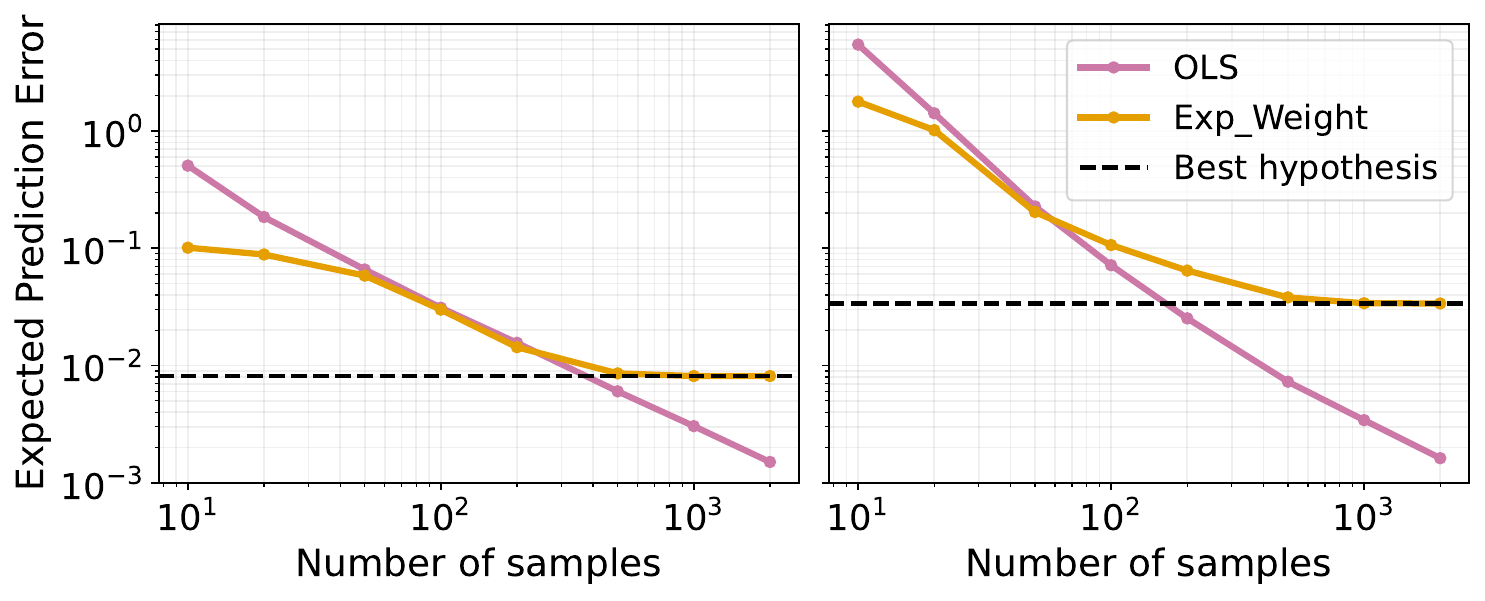}
    \caption{Excess risk for our estimator and \ac{OLS} for i.i.d. data (left) and trajectory data (right).}
    \label{fig:risk}
\end{figure}

To compare the performance of \ac{OLS} and our algorithm we consider the expected prediction error of the estimate $\hat \theta$ which is defined by
\begin{equation}
\Expect_{\theta_*}\left[\mathrm{PredErr}(\hat \theta) \right]\coloneqq \Expect_{\theta_*} \left[\sum_{t=0}^{T-1}\Vert (\hat \theta - \theta)z(t)\Vert^2 \right]. 
\end{equation}  
In Figure~\ref{fig:risk} we display the expected prediction error for different number of samples.
It can be observed that, in low data regimes, our algorithm, which uses the finite hypothesis class, outperforms \ac{OLS}. 
In high data regimes, where our algorithm has converged to $\hat \Theta_{*,T}$, \ac{OLS} achieves a better expected prediction error since \ac{OLS} is not restricted to the misspecified hypothesis class. 
While this is expected as discussed in Section~\ref{sec:discussion}, this yields the question, whether an adaptive sampling strategy which sequentially adds and removes candidates from $\SSet$ could yield faster performance compared to \ac{OLS}. We see a theoretical investigation of this topic as an interesting direction for future research.

\section{Conclusion}
In this work we consider the problem of identification from a finite hypothesis class under model misspecification. 
We provide lower bounds on the failure probability any algorithm must satisfy and provide an exponential weighting-based algorithm that achieves the lower bound up to a multiplicative factor. We suspect that this mismatch is an artifact of our proof of the  lower bound.
Our work shows that identifying an unknown dynamical system from a finite hypothesis class does still work even in presence of misspecifications, and opens up several implications and new research direction. 
First, in light of the dependence of the estimation target on the data collection scheme it is an interesting topic to understand how experiment design techniques should be applied in the misspecified setting. 
Second, we see the development of algorithms that sequentially adapt the hypothesis class as a potential avenue to obtain fast identification in low data regimes.

\appendix
\section{Supplementary Results}

\begin{prop}\label{prop:SubExp_new}
    Consider the random Gaussian vector $z\sim \N(0, I)$ of dimension $n_z$ and a symmetric matrix $\Xi\in \mathbb{R}^{n_z\times n_z}$. Let $\eta\le \frac{1}{4 \rho(\Xi)}$. Then 
    \begin{equation}\label{eq:subExp_new}
        \Expect \left[e^{-\eta z^\top \Xi z}\right] \le \exp\left(-\eta \left(\tr {\Xi}-2\eta\Vert \Xi\Vert_\mathrm{F}^2\right)\right).
    \end{equation}
\end{prop}
\begin{pf}
    Denote $\lambda_j = \lambda_j(\Xi)$ the $j$-th eigenvalue of $\Xi$. 
    Since $\Xi$ is real and symmetric there exists an orthogonal matrix $U$ such that $M = U \mathrm{diag}(\lambda_1, \dots, \lambda_{n_z}) U^\top$. 
    Hence we can write the quadratic form as 
    \begin{align*}
        z^\top M z &= z^\top U \begin{bmatrix}
        \lambda_1 & & \\ & \ddots & \\ & & \lambda_{n_z}
        \end{bmatrix} U^\top z = \zeta^\top \begin{bmatrix}
        \lambda_1 & & \\ & \ddots & \\ & & \lambda_{n_z}
        \end{bmatrix} \zeta\\ 
        &= \sum_{j=1}^{n_z} \zeta_j^2 \lambda_j
    \end{align*}
    where $\zeta = U^\top z \simiid\mathcal{N}(0, I)$ and hence $\zeta_i \simiid \N(0, 1)$. 
    Using the Gaussian integral for scalar normal random variables we obtain that as long as $1+ 2 \eta \lambda_j >0$ (which is implied by $\eta\le \frac{1}{4 \rho(\Xi)}$) it holds that 
    \begin{equation}\label{eq:proofAppendixInterm}
        \Expect \left[e^{-\eta z^\top \Xi z}\right] = \prod_{j=1}^{n_z} (1+2\eta\lambda_j)^{-\frac{1}{2}}.
    \end{equation}
    Note that $\log(1+x) \ge x -x^2$ as long as $x \ge -\frac12$. Thus, after taking the logarithm of \eqref{eq:proofAppendixInterm} and invoking the condition on $\eta$ we obtain
    \begin{align}
        \log(\prod_{j=1}^{n_z} (1+2\eta\lambda_j)^{-\frac{1}{2}}) &= \sum_{j=1}^{n_z} -\frac{1}{2} \log(1+2\eta\lambda_j) \\
        &\le -\sum_{j=1}^{n_z} \eta \lambda_j - 2 \eta^2 \lambda_j^2\\
        &= - \eta \tr{\Xi} + 2 \eta^2 \Vert \Xi \Vert_\mathrm{F}^2.
    \end{align}
    Taking the exponential completes the proof.
\end{pf}

\begin{prop}\label{prop:derivXi}
    There exists a symmetric matrix $\Xi$ so that 
    \begin{equation}
        \varepsilon_{\theta_i}(T) - \varepsilon_{\hat \theta_*}(T) \le \zeta^\top \Xi \zeta, \qquad \text{where $\zeta\sim \N(0, I)$.}
    \end{equation}
    Further, $\Vert \Xi \Vert_\mathrm{F}^2 \le T n_z \bar \rho_{\theta_i}$,
    where $\bar \rho_{\theta_i}$ is defined as follows
    \begin{itemize}
        \item Single trajectory data or data from multiple independent trajectories: Define $\ell \coloneqq 1 + \sqrt{1+\Vert B_*\Vert^2}\gamma(A_*)$, $\gamma(A_*) \coloneqq \sum_{s=1}^{T-1} \Vert A_*^s\Vert$ and $\bar \sigma \coloneqq  \max\{\Vert \Sigma_u \Vert, \sigma_w^2\}$, then 
        \begin{equation}\label{eq:barRho_traj}
            \bar \rho_{\theta_i} = \bar \sigma \left(\ell^2 \max \{\Vert \Delta\theta_i\Vert^2, \Vert \Delta \hat \theta_*\Vert^2\} + 2 \ell \Vert \Delta \hat \theta_i\Vert\right).
        \end{equation}
        \item Data sampled i.i.d.: Define $\bar \sigma \coloneqq \max \{\Vert \Sigma_x\Vert , \Vert \Sigma_u \Vert, \sigma_w^2\}$ then 
        \begin{equation}\label{eq:barRho_iid}
            \bar \rho_{\theta_i} = \bar \sigma\left(\max \{\Vert \Delta\theta_i\Vert^2, \Vert \Delta \hat \theta_*\Vert^2\} + 2 \Vert \Delta \hat \theta_i\Vert\right).
        \end{equation}
    \end{itemize}
\end{prop}
\begin{pf}
    First, we derive a matrix $\Xi$ such that we can rewrite~\eqref{eq:decompositionError} in the form $\zeta^\top \Xi \zeta$, where $\zeta \sim \N(0, I)$. We then provide the desired norm bound.    
    To this end, we consider each data collection scheme separately. \\
    \emph{Single trajectory data:} 
    Defining $v(t) \coloneqq \begin{bmatrix}
        u(t) ^\top w(t)^\top
    \end{bmatrix}^\top $ we obtain 
    \begin{equation}
        z(t) = \sum_{s=0}^{t-1} \begin{bmatrix}
            A_*^{t-1-s} B_* &  A_*^{t-1-s} \\ 0 & 0
        \end{bmatrix} v(s) + \begin{bmatrix}
            0 & 0 \\ 
            I_{n_u} & 0
        \end{bmatrix}
        v(t).
    \end{equation}
    Then using the stacked vector $V = \begin{bmatrix}
        v(0)^\top & \dots & v(T-1)^\top
    \end{bmatrix}^\top $ which satisfies 
    $V \sim \N(0, \Sigma_V)$ with 
    \begin{equation}
        \Sigma_V = I_T \otimes \begin{bmatrix}
            \Sigma_u & 0 \\ 0 & \sigma_w^2 I_{n_x}
        \end{bmatrix}.
    \end{equation}
    Thus for $Z = \begin{bmatrix}
        z(0)^\top & \dots & z(T-1)^\top
    \end{bmatrix}^\top$  we have that $Z = L_T V $, where $L_T$ is a lower triangular block-Toeplitz matrix. Similarly we can express $W= \begin{bmatrix}
        w(0)^\top & \dots & w(T-1)^\top
    \end{bmatrix}^\top$ as $W = E V$ where $\Vert E \Vert = 1$.
    Putting everything together and defining $\xi = \Sigma_V^{\frac{1}{2}}V \sim \N(0, I)$  we obtain 
    \begin{equation}
        \begin{aligned}
            \Xi=&\Sigma_V^{\frac{1}{2}}\Big(L_T^\top\big(I_T\otimes(\Delta\theta_i^\top\Delta\theta_i-\Delta\hat\theta_*^\top\Delta\hat\theta_*)\big)L_T\\
            &-L_T^\top(I_T\otimes \Delta \hat \theta_i)^\top E-E^\top(I_T\otimes \Delta \hat \theta_i)L_T\Big)\Sigma_V^{\frac{1}{2}}.
        \end{aligned}
    \end{equation}
    Using standard norm properties we obtain 
    \begin{equation}\label{eq:boundXiTraj}
         \Vert \Xi \Vert \le \Vert \Sigma_V \Vert \big( \Vert L_T \Vert^2 \max\{\Vert \Delta\theta_i\Vert^2, \Vert \Delta \hat \theta_*\Vert^2\} + 2 \Vert L_T \Vert \Vert \Delta \hat\theta_i\Vert \big).
    \end{equation}
    Note that since $\Sigma_V$ is block-diagonal we have 
    \begin{equation}
        \Vert \Sigma_V\Vert \le \max\{\Vert \Sigma_u \Vert, \sigma_w^2\}.
    \end{equation}
    Further we have 
    \begin{equation}
        \begin{aligned}
            \Vert L_T\Vert &\le 1 + \sqrt{1 + \Vert B_*\Vert^2} \sum_{s=1}^{T-1} \Vert A_*^s\Vert \\ 
            &\le 1 + \sqrt{1+\Vert B_*\Vert^2}\gamma(A_*).
        \end{aligned}
    \end{equation}
    Plugging these bounds into \eqref{eq:boundXiTraj} yields the result. 
    The result for multiple trajectories follows immediate by stacking them together and following the same steps. 
    \\
    \emph{Data sampled i.i.d.:} 
    Define 
    \begin{equation}
    V = \begin{bmatrix}
        z(0)^\top & w(0)^\top & \dots & z(T-1)^\top & w(T-1)^\top
    \end{bmatrix},
    \end{equation}
    which satisfies 
    $V \sim \N(0, \Sigma_V)$ with 
    \begin{equation}
        \Sigma_V = I_T \otimes \begin{bmatrix}
            \Sigma_z & 0 \\ 0 & \sigma_w^2 I_{n_x}
        \end{bmatrix}.
    \end{equation}
    It then follows that 
    \begin{equation}
            \begin{aligned}
                \Xi=&\Sigma_V^{\frac{1}{2}}\Big(L^\top\big(I_T\otimes(\Delta\theta_i^\top\Delta\theta_i-\Delta\hat\theta_*^\top\Delta\hat\theta_*)\big)L\\
                &-L^\top(I_T\otimes \Delta \hat \theta_i)^\top E-E^\top(I_T\otimes \Delta \hat \theta_i)L\Big)\Sigma_V^{\frac{1}{2}},
            \end{aligned}
    \end{equation}
    with $L = I_T\otimes \begin{bmatrix}
        I_{n_z} &0 \\ 0 &0 
    \end{bmatrix} $ and $E= I_T\otimes \begin{bmatrix}
        0 &0 \\ 0 & I_{n_x} 
    \end{bmatrix}$.
    The result follows by using standard properties of norms of matrices.
\end{pf}

\section*{DECLARATION OF GENERATIVE AI AND AI-ASSISTED TECHNOLOGIES IN THE WRITING PROCESS}
During the preparation of this work, the authors used Claude (claude.ai) to assist with proofreading, checking mathematical proofs, and developing code for the numerical evaluation. After using this tool/service, the author(s) reviewed and edited the content as needed and take(s) full responsibility for the content of the publication.

\bibliography{references}             


\end{document}